\documentclass[a4paper,11pt]{article}
\usepackage{pos}
\usepackage{caption}
\usepackage{subcaption}

\title{Central exclusive production of charged particle pairs in proton-proton collisions at $\sqrt{s}=200$~GeV with the STAR detector at RHIC}
\ShortTitle{Central exclusive production in $pp$ collisions at $\sqrt{s} = 200$~GeV with STAR}

\author*[\dag]{Rafal Sikora}
\author{~~for the STAR Collaboration}
\notes{\note{This work was partly supported by the National Science Centre of Poland under grant number
		UMO-2018/30/M/ST2/00395}}
\affiliation[]{AGH University of Science and Technology,\\
	Faculty of Physics and Applied Computer Science,\\
	Al. Mickiewicza 30, 30-055, Krakow, Poland}

\emailAdd{rafal.sikora@agh.edu.pl}

\abstract{The measurement of the central exclusive production of charged hadron pairs $h^{+}h^{-}$ ($h = \pi, K, p$) by the STAR experiment at RHIC is reported. The data from proton-proton collisions at $\sqrt{s} = 200$~GeV were used in this study. The pairs of charged hadrons produced in the reaction $pp\to p^\prime+h^{+}h^{-}+p^\prime$ were reconstructed from the tracks in the central detector, while the forward-scattered protons were measured in the Roman Pot system. Differential cross sections were measured in the fiducial region determined by the geometrical acceptance of the experimental setup. They were compared to phenomenological predictions based on the Double Pomeron Exchange model. The fiducial $\pi^+\pi^-$ cross section was extrapolated to the Lorentz-invariant region, which allowed decomposition of the invariant mass spectrum into continuum and resonant contributions.
}

\FullConference{%
  40th International Conference on High Energy physics - ICHEP2020\\
  July 28 - August 6, 2020\\
  Prague, Czech Republic (virtual meeting)
}

\begin{document}
\maketitle

\section{Introduction}

The STAR experiment~\cite{STAR} at RHIC reports a high-statistics measurement of the Central Exclusive Production (CEP) process in proton-proton collisions, $p+p\rightarrow p'+X+p'$, based on the data collected at the center-of-mass energy $\sqrt{s}=200$~GeV~\cite{cepSTAR}. The process is recognized when all particles of the centrally-produced neutral state $X$ (here denoting a pair of opposite-charge hadrons, $h^+h^-$) are well separated in the rapidity space from the intact scattered beam (or target) particles ($p'$). In the limit of high center-of-mass energies the process occurs mainly through the Double Pomeron Exchange (DPE) mechanism. The QCD representation of the Pomeron as a pair of gluons in the color singlet makes the process potentially suitable for the production of the gluon bound states (glueballs). However, in this process also regular mesons (resonances) are produced, accompanied by the direct hadron pair production, all interfering with each other. On top of this some additional soft exchanges are possible between the particles involved in the process, spoiling the exclusivity of the reaction (absorption effects). Theoretical complexity of this topologically-simple process makes it an important object of experimental and theoretical studies~\cite{LS,Durham,MBR}.

\section{Event selection}
\begin{figure}[b!]
	\centering
	\parbox{0.475\textwidth}{
		\centering
		\begin{subfigure}[b]{\linewidth}{
				\subcaptionbox{\vspace*{-0pt}\label{fig:rp_hits_2d}}{\includegraphics[width=\linewidth]{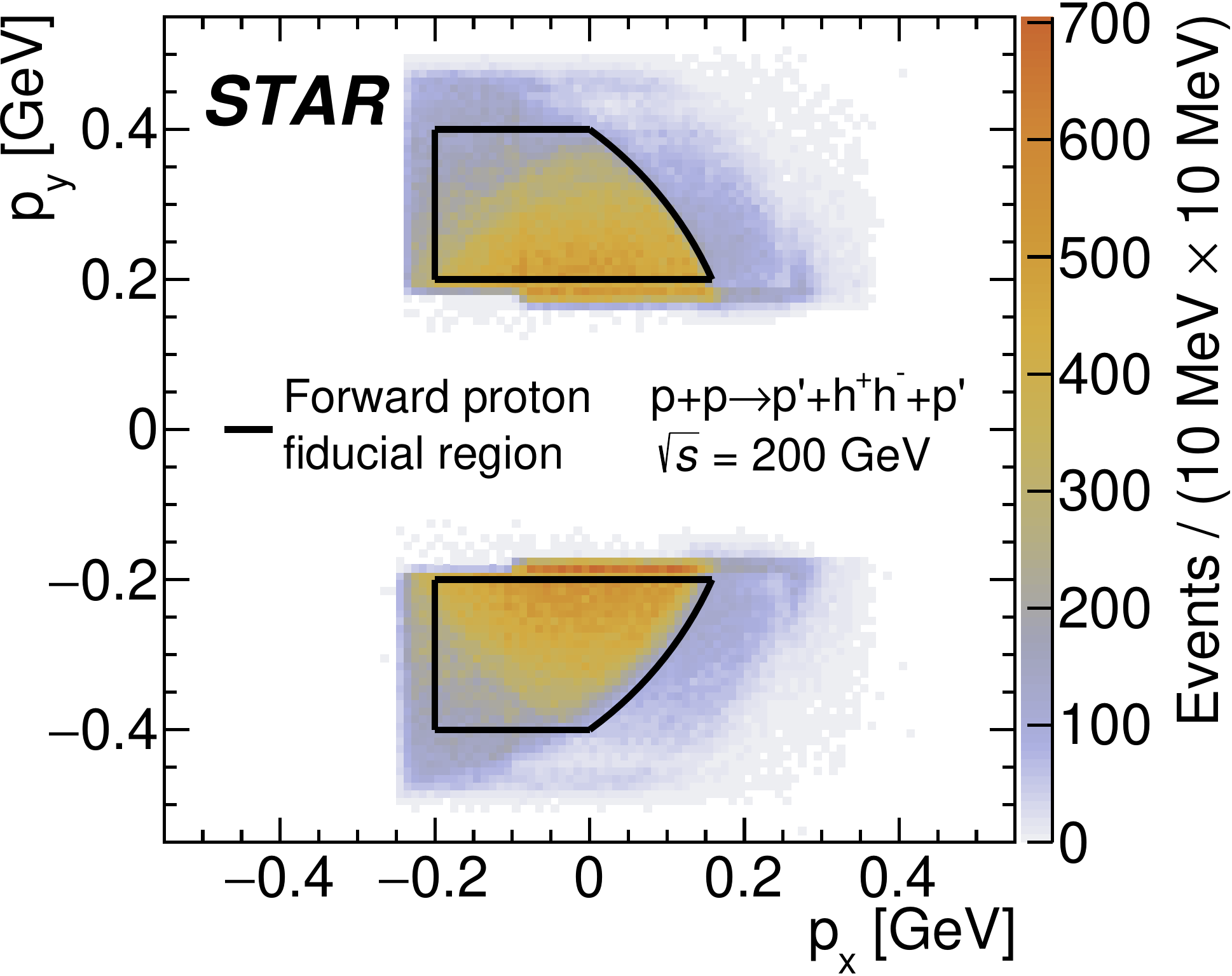}\vspace{-10pt}}}\vspace{-7pt}
		\end{subfigure}
	}%
	\quad
	\parbox{0.475\textwidth}{%
		\centering
		\begin{subfigure}[b]{\linewidth}{\vspace*{-5pt}
				\subcaptionbox{\vspace*{-0pt}\label{fig:rp_t}}{\includegraphics[width=\linewidth]{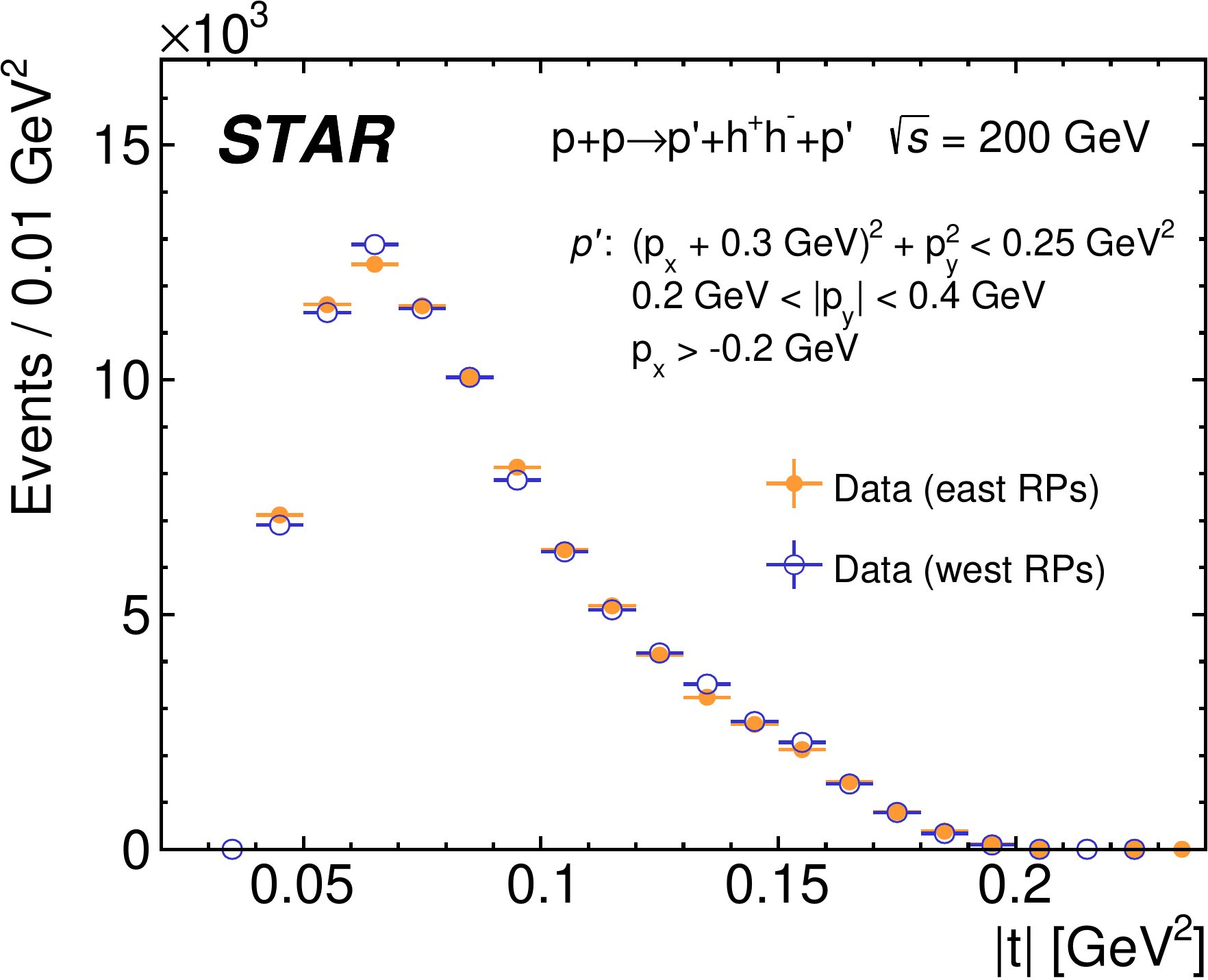}\vspace{-10pt}}}\vspace{-7pt}
		\end{subfigure}
	}%
	\caption[Merged distributions of diffractively scattered protons momenta $p_y$ vs. $p_x$ and squared four momenta transfers at the proton vertices in exclusive $h^{+}h^{-}$ events.]{(\subref{fig:rp_hits_2d}) Merged distributions of diffractively scattered protons momenta $p_y$ vs. $p_x$ in exclusive $h^{+}h^{-}$ events reconstructed with the East and West RP stations, are shown together with the kinematic region used in the measurement marked with the black line. (\subref{fig:rp_t}) Distributions of measured squared four momenta transfers at the proton vertices for exclusive $h^{+}h^{-}$ events with all particles in the fiducial phase space are shown for East and West stations with yellow and blue colour, respectively.}
	\label{fig:rp_hits}
\end{figure}

The data used for the measurement were collected with a dedicated trigger, which required coincidence of the trigger signals in Roman Pot (RP) detectors~\cite{elasticSTAR} on two sides (East and West) of the interaction point (IP), accompanied by the trigger signal in the Time-of-Flight (TOF) system ($|\eta|<0.9$) and a veto on activity in the Beam-Beam Counter (BBC) scintillators ($3.3<|\eta|<5$). In the offline analysis exactly two good-quality RP tracks were required, one per each side of the IP. The tracks of forward-scattered protons were required to lie within the fiducial region described as a function of transverse momentum components (Fig.~\ref{fig:rp_hits_2d}). As shown in Fig.~\ref{fig:rp_t}, the four-momentum transfers, $|t|$, of the forward-scattered protons from the fiducial region, were generally greater than $0.04$~GeV$^{2}$, which suppressed contributions from the photon-Pomeron exchanges.

The tracks of the centrally-produced $h^+h^-$ pairs were reconstructed in the Time Projection Chamber (TPC). Exactly two good-quality opposite-sign TPC tracks were required in the single primary vertex, both matched with the hits in the TOF detector. They were required to satisfy minimum transverse momentum cut, $p_{\text{T}}>0.2$~GeV, and obey pseudorapidity constraint $|\eta|<0.7$. Using the specific energy loss reconstructed for each track, $dE/dx$, and the time of flight from the vertex to the TOF detector, it was possible to identify $\pi^+\pi^-$, $K^+K^-$ and $p\bar{p}$ pairs.

\begin{figure}[hb]
	\centering
	\parbox{0.485\textwidth}{
		\centering
		\begin{subfigure}[b]{\linewidth}{
				\subcaptionbox{\label{fig:missingPt}}{\includegraphics[width=\linewidth]{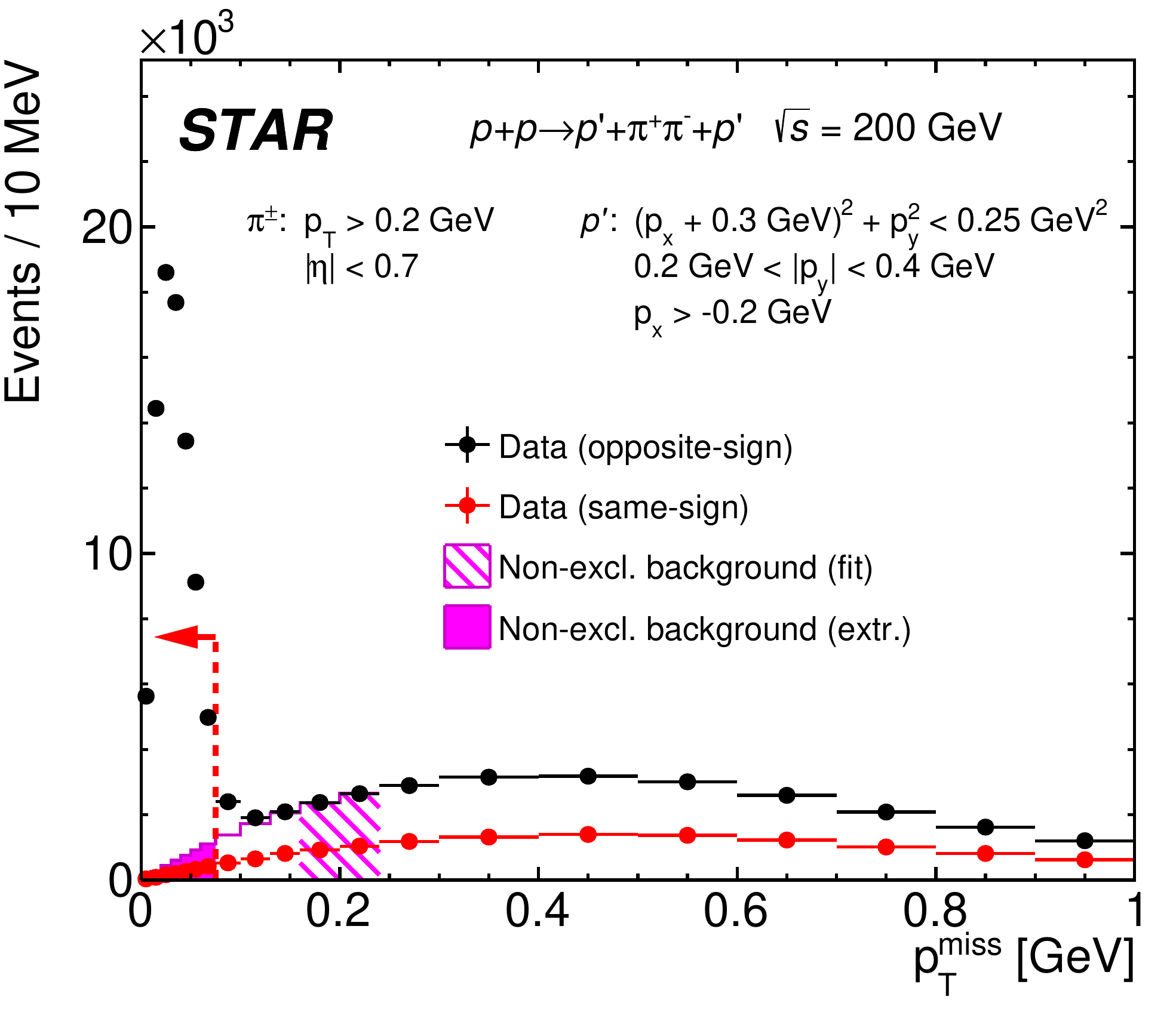}\vspace{-10pt}}}\vspace{-7pt}
		\end{subfigure}
	}%
	\quad%
	\parbox{0.485\textwidth}{%
		\centering
		\begin{subfigure}[b]{\linewidth}{
				\subcaptionbox{\label{fig:invMass}}{\includegraphics[width=\linewidth]{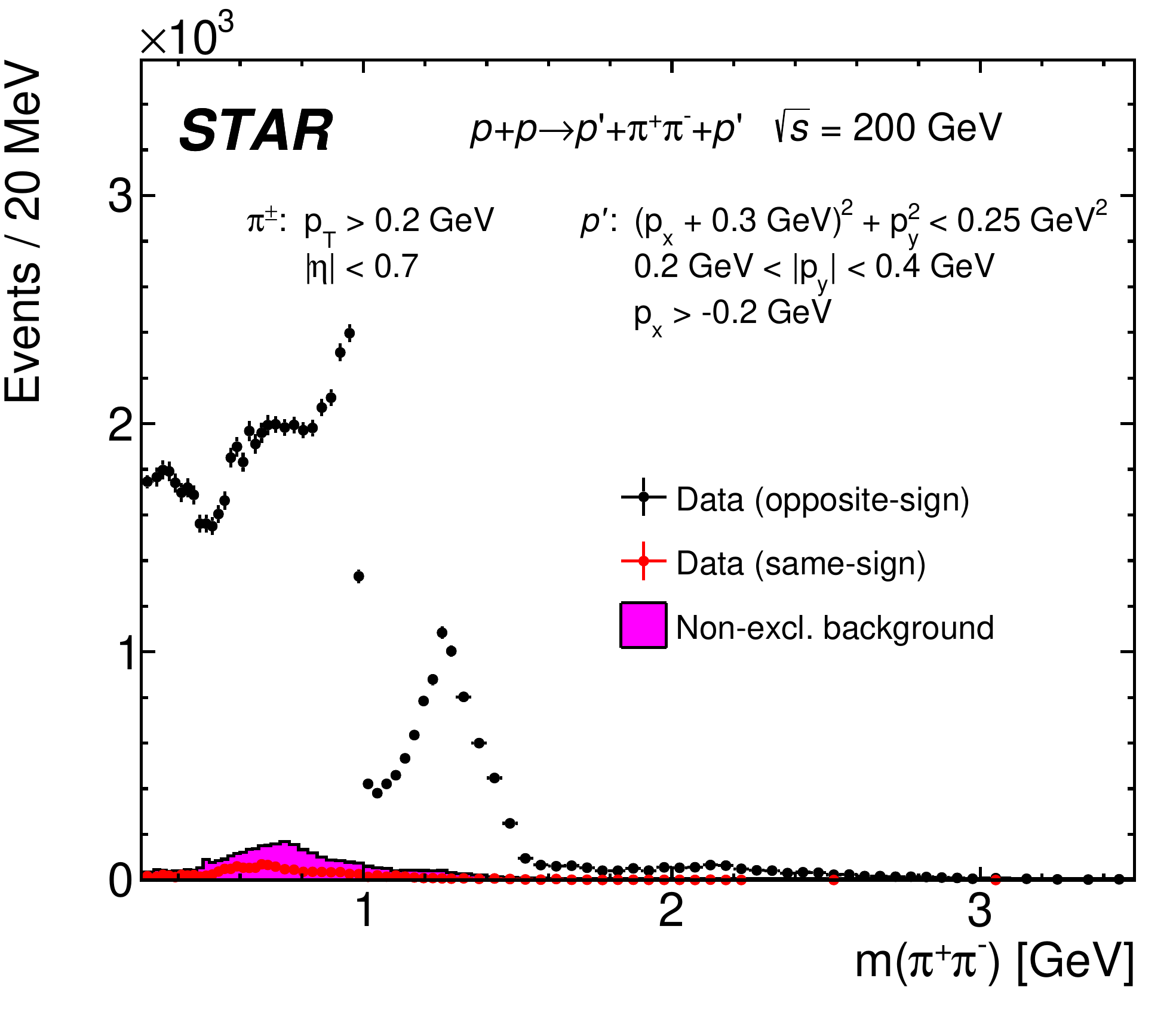}\vspace{-10pt}}}\vspace{-7pt}
		\end{subfigure}
	}%
	\caption{Uncorrected distributions of the CEP event candidates' (\subref{fig:missingPt}) missing transverse momentum $p_\mathrm{T}^\mathrm{\scriptstyle miss}$ and (\subref{fig:invMass}) invariant mass of the charged particle pairs produced in the final state for $\pi^+\pi^-$ pairs. Invariant mass distribution is obtained for the signal dominated region marked with the red arrows on the $p_\mathrm{T}^\mathrm{\scriptstyle miss}$ plot. Distributions for opposite-sign and same-sign particle pairs are shown as black and red symbols, respectively. The vertical error bars represent statistical uncertainties. The horizontal bars represent bin sizes. Solid magenta histograms correspond to the estimated non-exclusive background, determined differentially from the number of counts in the hatched range $0.16\,\text{GeV}<p_\mathrm{T}^\mathrm{\scriptstyle miss}<0.24\,\text{GeV}$, and extrapolated to the signal region indicated with dashed red line and arrow.}
	\label{fig:missingPt_invMass}\vspace*{-10pt}
\end{figure}

The above selections were supplemented with a few additional criteria. The most important requirement utilized the fact that in addition to the central state particles also the forward-going protons were measured. It was possible to construct the total transverse momentum of all four final-state particles, $p_\mathrm{T}^\mathrm{\scriptstyle miss}$, which efficiently verified exclusivity of an event. As shown in Fig.~\ref{fig:missingPt}, a cut $p_\mathrm{T}^\mathrm{\scriptstyle miss}<75$~MeV was used to select CEP events represented by a narrow peak near $p_\mathrm{T}^\mathrm{\scriptstyle miss}=0$. The width of the peak was determined by the transverse momentum resolutions dominated by the angular divergence of the proton beams. The data-driven method based on the $p_\mathrm{T}^\mathrm{\scriptstyle miss}$ distribution was used to estimate the non-exclusive background, whose amount in the final samples was suppressed to the level of several percent. The method was also used differentially as a function of the physics observables, e.g. invariant mass of the $\pi^+\pi^-$ pair, as shown in Fig.~\ref{fig:invMass}.

\section{Results}

\begin{figure}[b!]
	\centering
	\parbox{0.485\textwidth}{
		\centering
		\begin{subfigure}[b]{0.7\linewidth}{
				\subcaptionbox{\vspace*{-0pt}\label{fig:deltaPhi}}{\includegraphics[width=\linewidth]{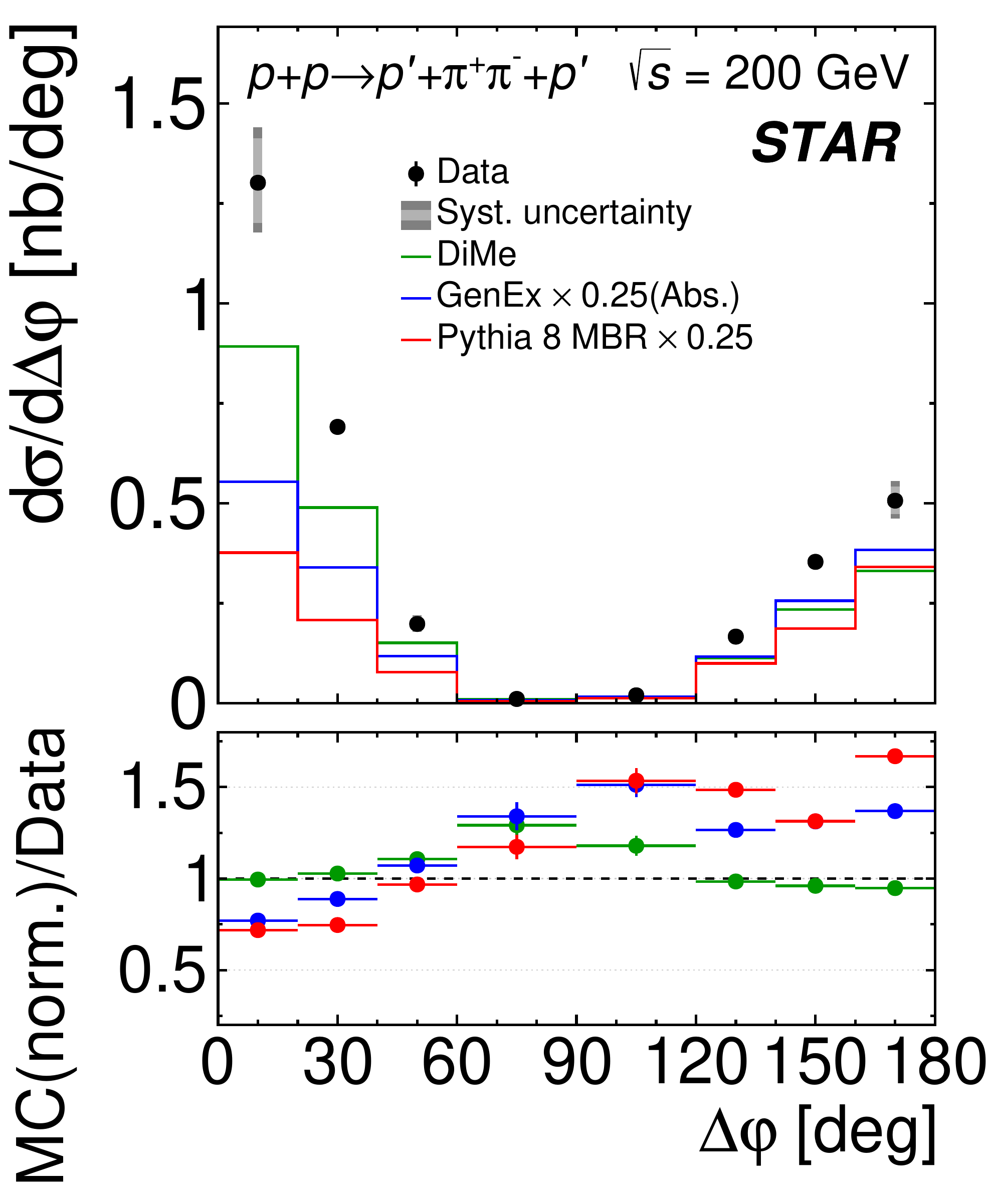}\vspace{-10pt}}}\vspace{-7pt}
		\end{subfigure}
	}%
	\quad%
	\parbox{0.485\textwidth}{%
		\centering
		\begin{subfigure}[b]{0.7\linewidth}{
				\subcaptionbox{\vspace*{-0pt}\label{fig:tSum}}{\includegraphics[width=\linewidth]{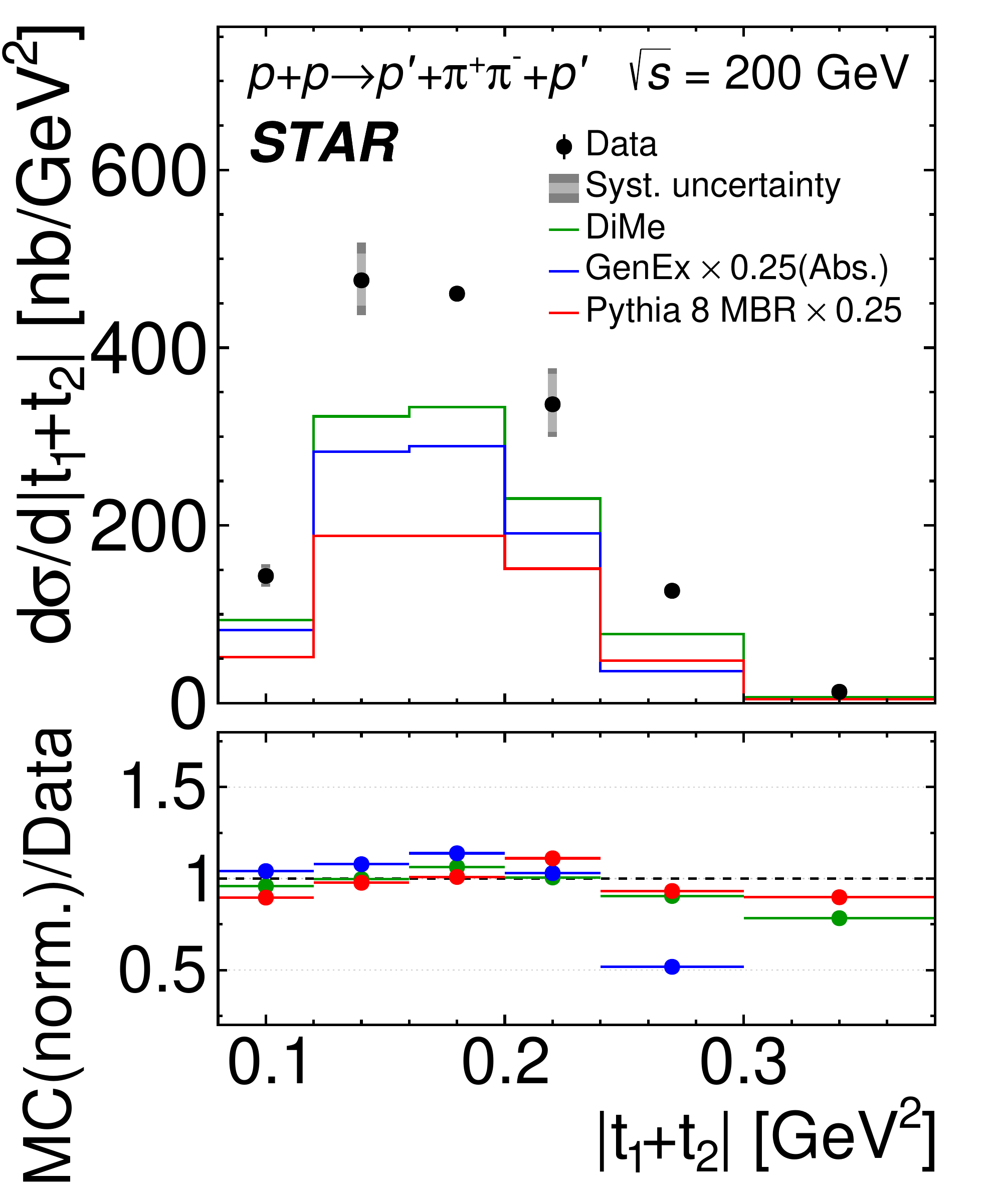}\vspace{-10pt}}}\vspace{-7pt}
		\end{subfigure}
	}%
	\caption{Differential fiducial cross sections for CEP of $\pi^+\pi^-$ pairs as a function of (\subref{fig:invMass_2pi_STAR_DeltaPhiLess90}) the difference of azimuthal angles of the forward-scattered protons, and (\subref{fig:invMass_2pi_STAR_DeltaPhiMore90}) the sum of the squares of the four-momenta transfers in the proton vertices. Data are shown as solid points with error bars representing the statistical uncertainties. The typical systematic uncertainties are shown as grey boxes for only few data points as they are almost fully correlated between neighbouring bins. Predictions from MC models GenEx~\cite{LS}, DiMe~\cite{Durham} and MBR~\cite{MBR} are shown as histograms.}
	\label{fig:deltaPhi_tSum}\vspace{15pt}
	\centering
	\parbox{0.485\textwidth}{
		\centering
		\begin{subfigure}[b]{\linewidth}{
				\subcaptionbox{\vspace*{-0pt}\label{fig:invMass_2pi_STAR_DeltaPhiLess90}}{\includegraphics[width=\linewidth]{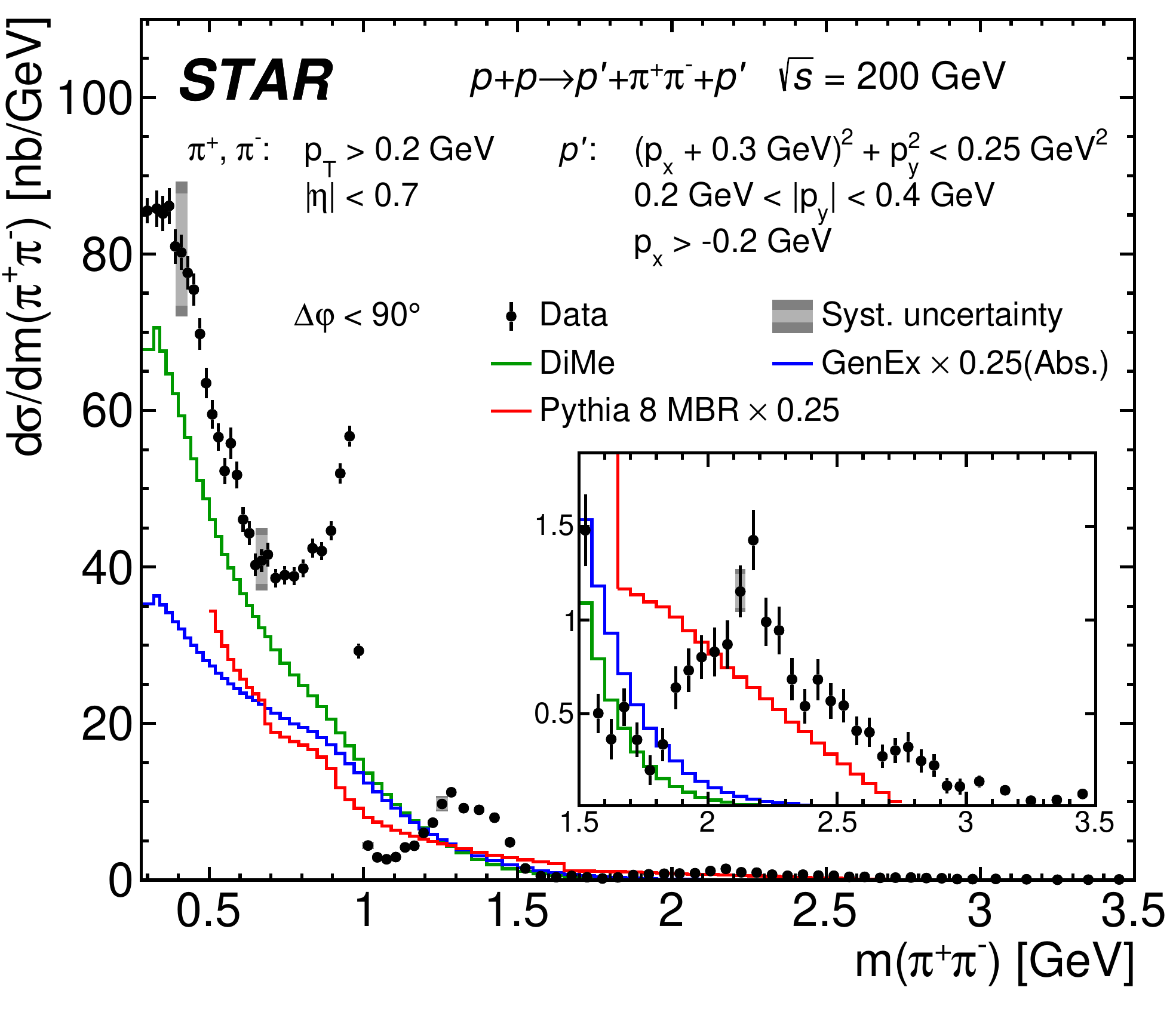}\vspace{-10pt}}}\vspace{-7pt}
		\end{subfigure}
	}%
	\quad%
	\parbox{0.485\textwidth}{%
		\centering
		\begin{subfigure}[b]{\linewidth}{
				\subcaptionbox{\vspace*{-0pt}\label{fig:invMass_2pi_STAR_DeltaPhiMore90}}{\includegraphics[width=\linewidth]{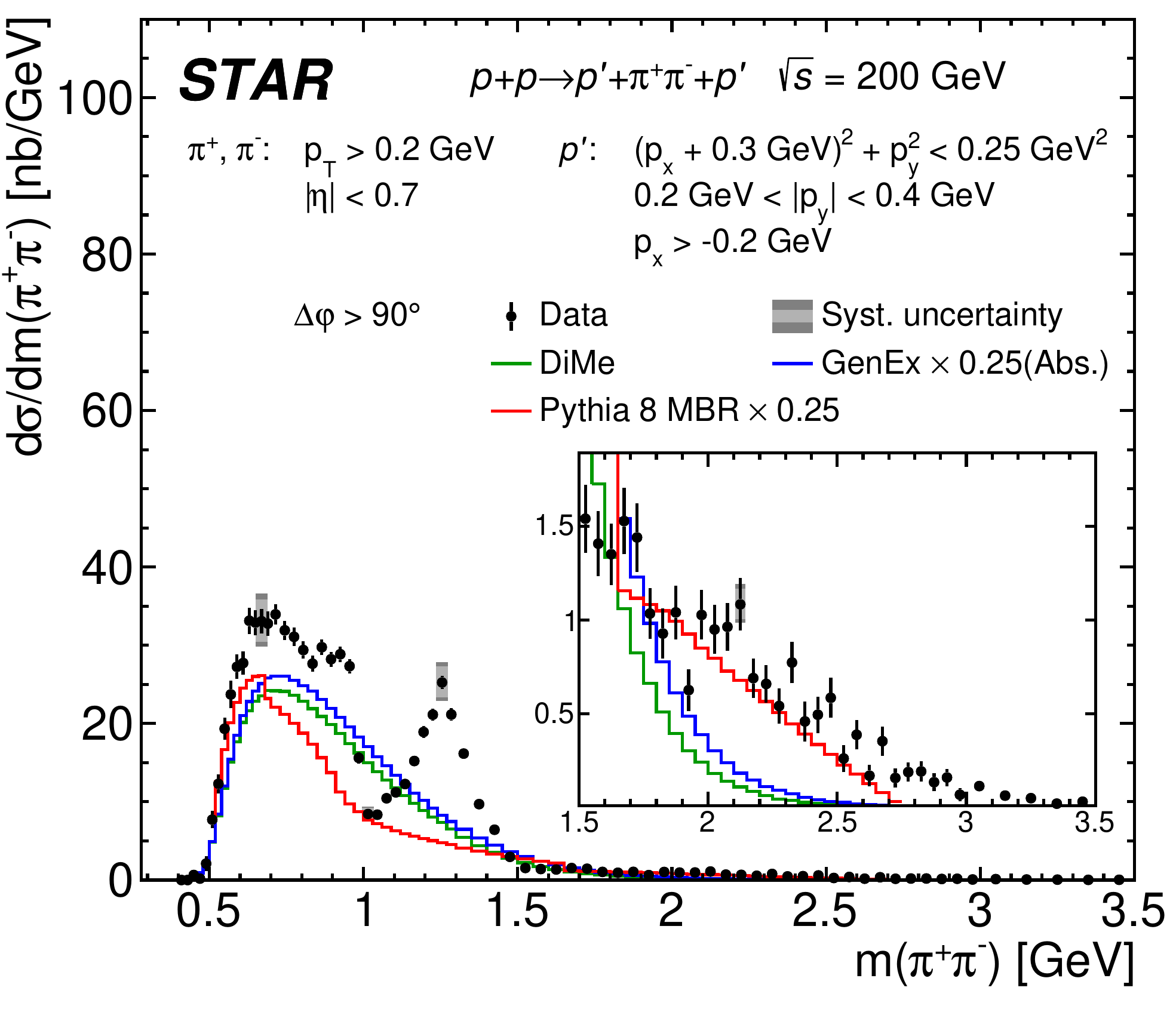}\vspace{-10pt}}}\vspace{-7pt}
		\end{subfigure}
	}
	\caption[Differential cross sections for CEP of $\pi^+\pi^-$ pairs as a function of the invariant mass of the pair in two $\Delta\upvarphi$ regions: $\Delta\upvarphi<90^\circ$ and $\Delta\upvarphi>90$ degree, measured in the fiducial region explained on the plots.]{Differential cross sections for CEP of $\pi^+\pi^-$ pairs as a function of the invariant mass of the pair in two $\Delta\upvarphi$ regions: (\subref{fig:invMass_2pi_STAR_DeltaPhiLess90}) $\Delta\upvarphi<90^\circ$ and (\subref{fig:invMass_2pi_STAR_DeltaPhiMore90}) $\Delta\upvarphi>90^\circ$, measured in the fiducial region explained on the plots. Data are shown as solid points with error bars representing the statistical uncertainties. The typical systematic uncertainties are shown as grey boxes for only few data points as they are almost fully correlated between neighbouring bins. Predictions from MC models GenEx, DiMe and MBR are shown as histograms.}
	\label{fig:invMass_2pi_STAR_DeltaPhi}
\end{figure}

\begin{figure}[t!]
	\centering
	\parbox{0.315\textwidth}{
		\centering
		\begin{subfigure}[b]{\linewidth}{
				\subcaptionbox{\vspace*{-2pt}\label{fig:Ratio_FinalResult_PhiCS_pion_MassBin_1}}{\includegraphics[width=\linewidth]{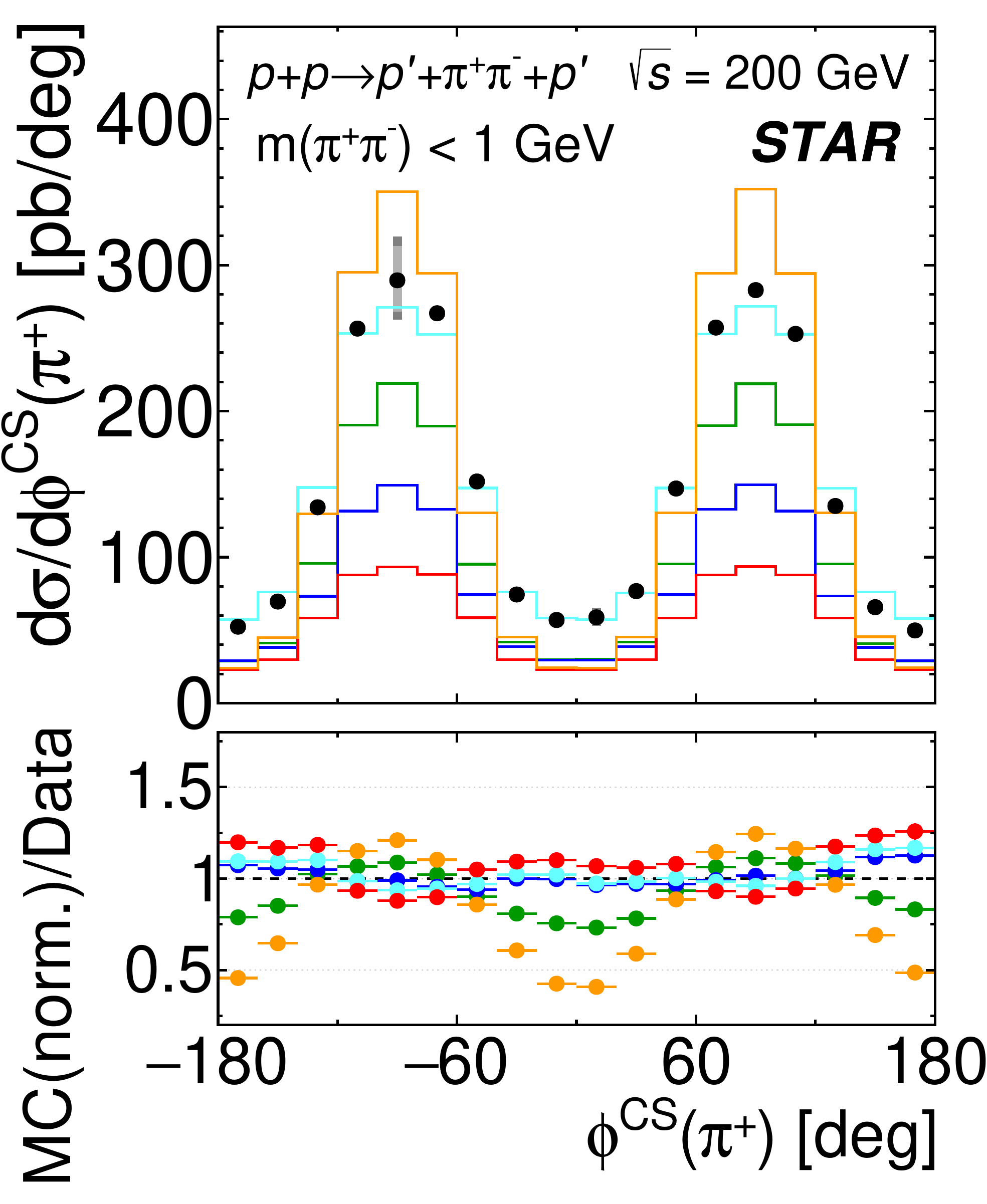}\vspace{-10pt}}}\vspace{-5pt}
		\end{subfigure}
	}%
	\quad%
	\parbox{0.315\textwidth}{%
		\centering
		\begin{subfigure}[b]{\linewidth}{
				\subcaptionbox{\vspace*{-2pt}\label{fig:Ratio_FinalResult_PhiCS_pion_MassBin_2}}{\includegraphics[width=\linewidth]{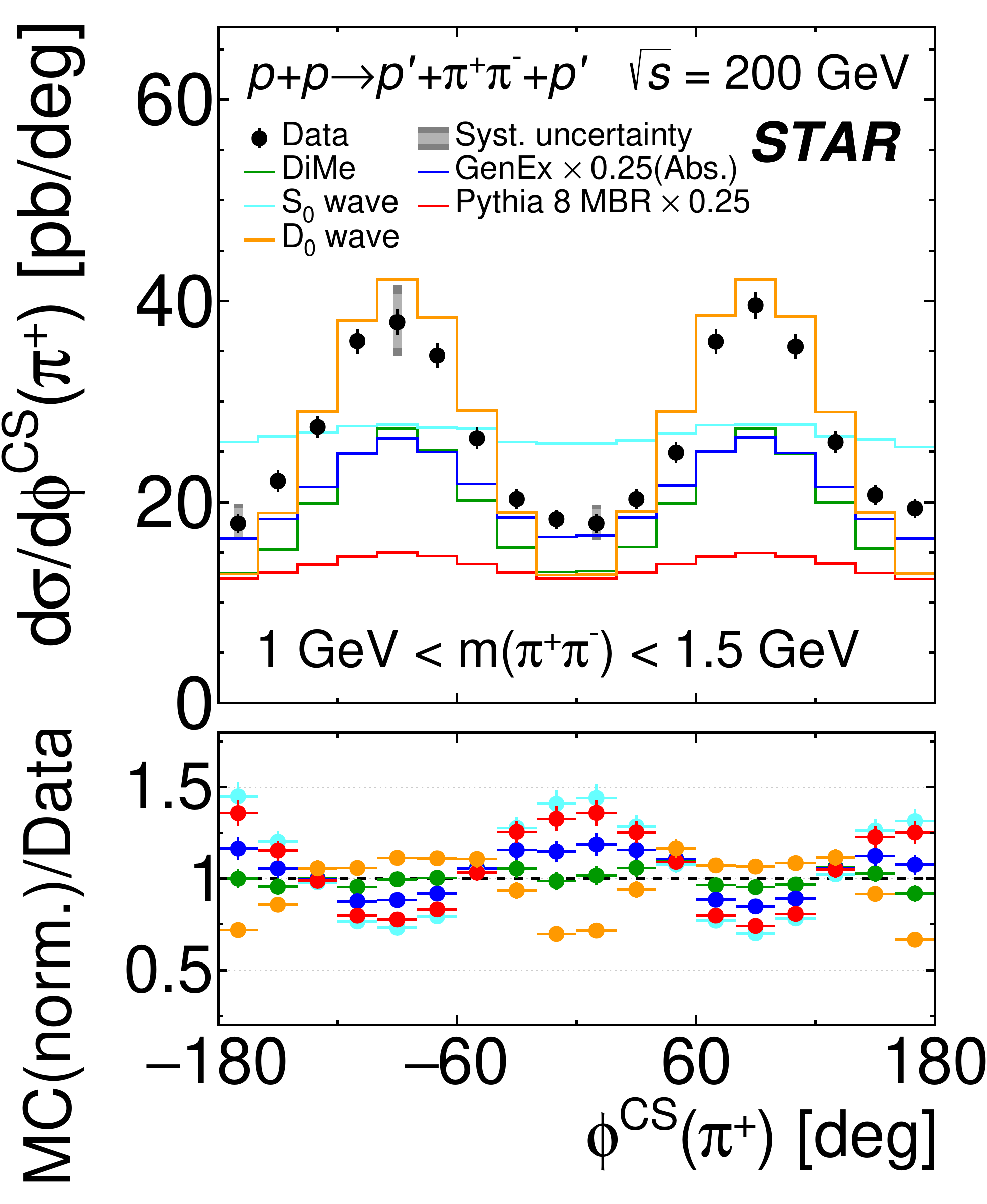}\vspace{-10pt}}}\vspace{-5pt}
		\end{subfigure}
	}%
	\quad%
		\parbox{0.315\textwidth}{%
		\centering
		\begin{subfigure}[b]{\linewidth}{
				\subcaptionbox{\vspace*{-2pt}\label{fig:Ratio_FinalResult_PhiCS_pion_MassBin_3}}{\includegraphics[width=\linewidth]{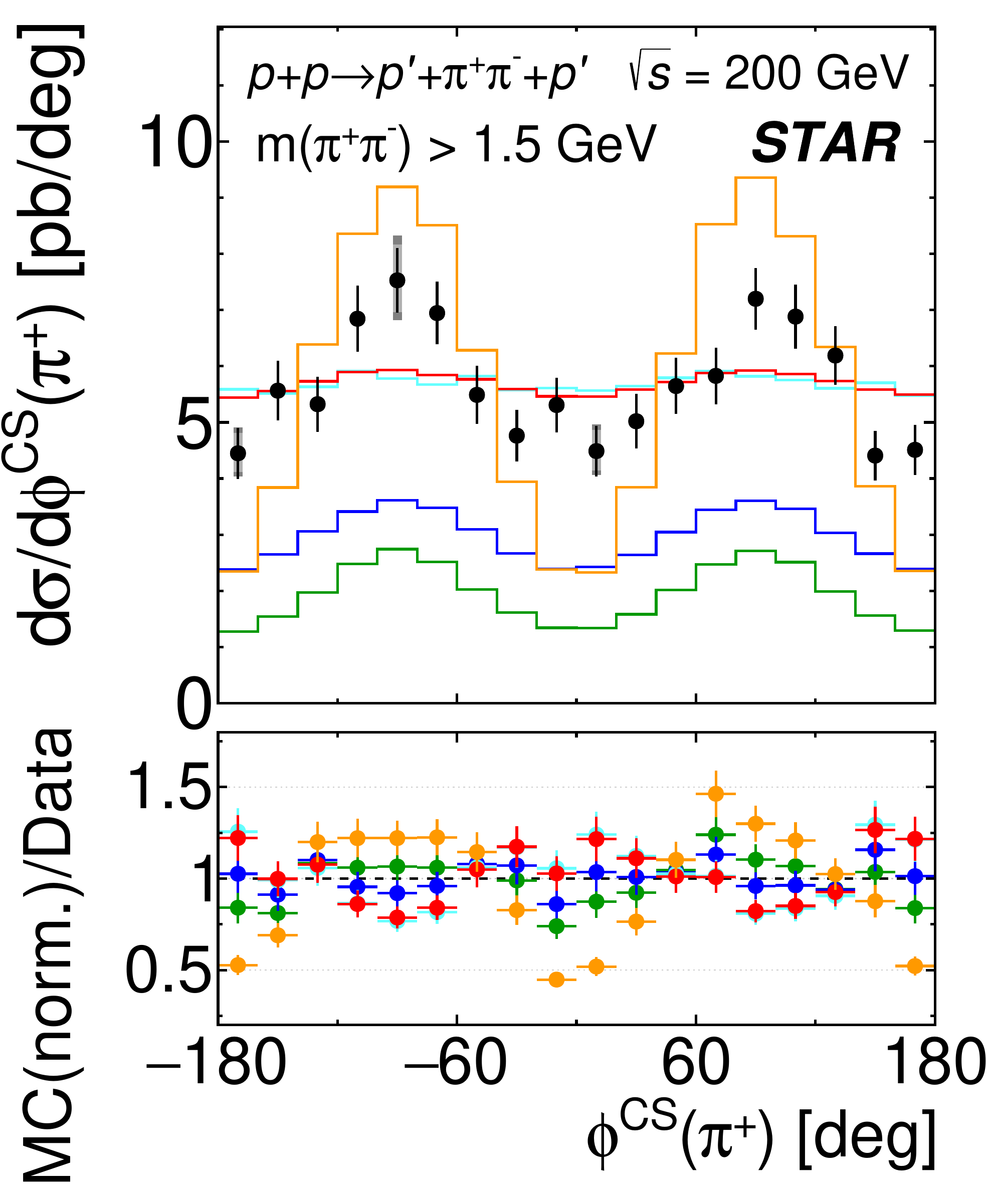}\vspace{-10pt}}}\vspace{-5pt}
		\end{subfigure}
	}
	\caption{Differential fiducial cross sections for CEP of $\pi^+\pi^-$ pairs as a function of $\upphi^\mathrm{CS}$ for three ranges of the $\pi^+\pi^-$ pair invariant mass: (\subref{fig:Ratio_FinalResult_PhiCS_pion_MassBin_1}) $m<1$~GeV, (\subref{fig:Ratio_FinalResult_PhiCS_pion_MassBin_2}) $1~\textrm{GeV}<m<1.5$~GeV and (\subref{fig:Ratio_FinalResult_PhiCS_pion_MassBin_3}) $m>1.5$~GeV. Data are shown as solid points with error bars representing the statistical uncertainties. The typical systematic uncertainties are shown as grey boxes for only few data points as they are almost fully correlated between neighbouring bins. Predictions from MC models GenEx, DiMe and MBR, as well as predictions for pure $S_0$ and $D_0$ waves, are shown as histograms. In the bottom panels, the ratios of the MC predictions scaled to data and the data are shown.}
	\label{fig:phiCS_MassBins_2Pi}
\end{figure}

\begin{figure}[t!] 
	\centering
	\includegraphics[width=\textwidth,page=1]{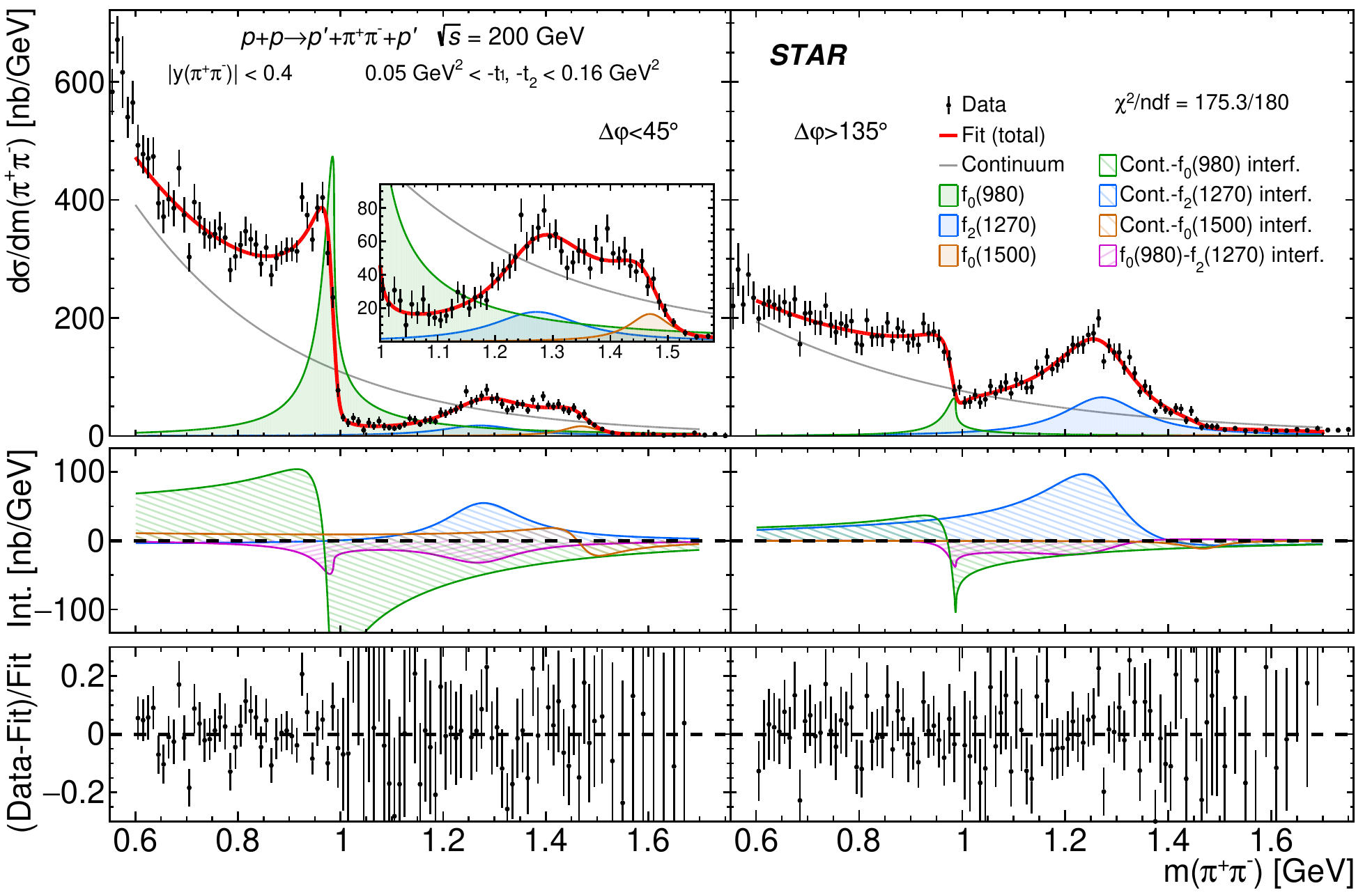}
	\caption[Differential cross section $d\sigma/dm(\pi^{+}\pi^{-})$ extrapolated from the fiducial region to $|y(\pi^{+}\pi^{-})|<0.4$ and $0.05~\text{GeV}^{2} < -t_{1}, -t_{2} < 0.16~\text{GeV}^{2}$]{Differential cross section $d\sigma/dm(\pi^{+}\pi^{-})$ extrapolated from the fiducial region to the Lorentz-invariant phase space given by the central-state rapidity, $|y(\pi^{+}\pi^{-})|<0.4$, and squared four-momentum transferred in forward proton vertices, $0.05\,\text{GeV}^{2} < -t_1, -t_2 < 0.16\,\text{GeV}^{2}$. The left and right columns show the cross sections for $\Delta\upvarphi<45^\circ$ and $\Delta\upvarphi>135^\circ$, respectively. The data are shown as black points with error bars representing statistical uncertainties. The result of the fit is drawn with a solid red line. The squared amplitudes for the continuum and resonance production are drawn with lines of different colors, as explained in the legend. The most significant interference terms are plotted in the middle panels, while the relative differences between each data point and the fitted model is shown in the bottom panels.}
	\label{fig:invMassFit}
\end{figure}

The primary results of this analysis are differential fiducial cross sections for CEP of $\pi^+\pi^-$, $K^+K^-$ and $p\bar{p}$ pairs. Figures~\ref{fig:deltaPhi} and~\ref{fig:tSum} show the differential fiducial cross sections for CEP of $\pi^+\pi^-$ pairs as a function of the azimuthal separation, $\Delta\upvarphi$, and the sum of the squared four-momentum transfers, $|t_1+t_2|$, of the forward-scattered protons, respectively. These observables represent the forward-scattered protons and are particularly sensitive to the absorption effects. Properties of the central hadron pairs were studied with respect to the configuration of the forward protons, which reflects the dynamics of the Pomeron exchange. Figure~\ref{fig:invMass_2pi_STAR_DeltaPhi} shows the invariant mass of the $\pi^+\pi^-$ pairs in two characteristic ranges of $\Delta\upvarphi$ angle. Many structures are found in the spectrum whose magnitudes significantly vary with the $\Delta\upvarphi$ angle. Studies of angular distributions, e.g. of the azimuthal angle of positively charged pion in the Collins-Soper frame~\cite{cs}, $\upphi^{\textrm{CS}}(\pi^+)$ (Fig.~\ref{fig:phiCS_MassBins_2Pi}), reveal that $\pi^+\pi^-$ pairs of the invariant mass below 1~GeV are dominated by the S-wave, while in the region between 1-1.5~GeV have significant contribution from the D-wave. A successful attempt to model the extrapolated invariant mass spectrum of $\pi^+\pi^-$ pairs was performed (Fig.~\ref{fig:invMassFit}), which confirms production, in addition to the continuum, of $f_0(980)$, $f_2(1270)$ and $f_0(1500)$. The last resonance, found to be dominantly produced at $\Delta\upvarphi<90^\circ$, is generally considered as a potential state with some gluon admixture.

\section{Summary}
Recent results on CEP of charged particle pairs in proton-proton collisions at $\sqrt{s}=200$~GeV obtained by the STAR experiment have been presented. It is currently the highest center-of-mass energy, at which this process was measured with detection of the forward-scattered particles. The data should help in understanding of the DPE mechanism and developing models of the process.

\vspace*{-8pt}
\bibliographystyle{JHEP}
\bibliography{references}

\providecommand{\href}[2]{#2}\begingroup\raggedright\begin{thebibliography}{1}

\bibitem{STAR}
{\scshape STAR} Collaboration, K.~Ackermann et~al., \emph{{STAR detector
  overview}},
  \href{http://dx.doi.org/10.1016/S0168-9002(02)01960-5}{\emph{Nucl. Instrum.
  Meth. A} {\bfseries 499} (2003) 624--632}.

\bibitem{cepSTAR}
{\scshape STAR} Collaboration, J.~Adam et~al., \emph{{Measurement of the
  central exclusive production of charged particle pairs in proton-proton
  collisions at $\sqrt{s} = 200$ GeV with the STAR detector at RHIC}},
  \href{http://dx.doi.org/10.1007/JHEP07(2020)178}{\emph{JHEP} {\bfseries 07}
  (2020) 178}, [\href{https://arxiv.org/abs/2004.11078}{{\ttfamily
  arXiv:2004.11078}}].

\bibitem{LS}
P.~Lebiedowicz and A.~Szczurek, \emph{{Exclusive $pp \to pp \pi^{+}\pi^{-}$
  reaction: From the threshold to LHC}},
  \href{http://dx.doi.org/10.1103/PhysRevD.81.036003}{\emph{Phys. Rev. D}
  {\bfseries 81} (2010) 036003},
  [\href{https://arxiv.org/abs/0912.0190}{{\ttfamily arXiv:0912.0190}}].

\bibitem{Durham}
L.~Harland-Lang, V.~Khoze and M.~Ryskin, \emph{{Modelling exclusive meson pair
  production at hadron colliders}},
  \href{http://dx.doi.org/10.1140/epjc/s10052-014-2848-9}{\emph{Eur. Phys. J.
  C} {\bfseries 74} (2014) 2848},
  [\href{https://arxiv.org/abs/1312.4553}{{\ttfamily arXiv:1312.4553}}].

\bibitem{MBR}
R.~Ciesielski and K.~Goulianos, \emph{{MBR Monte Carlo Simulation in PYTHIA8}},
  \href{http://dx.doi.org/10.22323/1.174.0301}{\emph{PoS} {\bfseries ICHEP2012}
  (2013) 301}, [\href{https://arxiv.org/abs/1205.1446}{{\ttfamily
  arXiv:1205.1446}}].

\bibitem{elasticSTAR}
{\scshape STAR} Collaboration, J.~Adam et~al., \emph{{Results on total and
  elastic cross sections in proton\textendash{}proton collisions at $\sqrt{s}$
  = 200 GeV}},
  \href{http://dx.doi.org/10.1016/j.physletb.2020.135663}{\emph{Phys. Lett. B}
  {\bfseries 808} (2020) 135663},
  [\href{https://arxiv.org/abs/2003.12136}{{\ttfamily arXiv:2003.12136}}].

\bibitem{cs}
J.~C. Collins and D.~E. Soper, \emph{{Angular Distribution of Dileptons in
  High-Energy Hadron Collisions}},
  \href{http://dx.doi.org/10.1103/PhysRevD.16.2219}{\emph{Phys. Rev. D}
  {\bfseries 16} (1977) 2219}.

\end{thebibliography}\endgroup

\end{document}